\documentclass[aps,prd,prabib,twocolumn,showpacs,nofootinbib]{revtex4}
\usepackage{graphicx} \usepackage{amsmath} \usepackage{amssymb}
\usepackage{amsfonts} \usepackage{bm}

\begin{document}

\newcommand{\be}{\begin{equation}} \newcommand{\ee}{\end{equation}}
\newcommand{\bea}{\begin{eqnarray}}\newcommand{\eea}{\end{eqnarray}}

\title{Non-Hermitian quantum mechanics in non-commutative space}

\author{Pulak Ranjan Giri} \email{pulakranjan.giri@saha.ac.in}

\affiliation{Theory Division, Saha Institute of Nuclear Physics, 1/AF
Bidhannagar, Calcutta 700064, India}

\author{P Roy} \email{pinaki@isical.ac.in}

\affiliation{Physics and applied mathematics Unit, Indian Statistical
Institute, kolkata 700108, India}

\begin{abstract}
We study non Hermitian quantum systems in noncommutative space as
well as a $\cal{PT}$-symmetric deformation of this space.
Specifically, a $\mathcal{PT}$-symmetric harmonic oscillator
together with  $iC(x_1+x_2)$ interaction is discussed in this space
and  solutions are obtained. It is shown that in the $\cal{PT}$
deformed noncommutative space the Hamiltonian may or may not possess
real eigenvalues depending on the choice of the noncommutative
parameters. However, it is shown that in standard noncommutative
space, the $iC(x_1+x_2)$ interaction generates only real eigenvalues
despite the fact that the Hamiltonian is not
$\mathcal{PT}$-symmetric. A complex interacting anisotropic
oscillator system has also been discussed.
\end{abstract}


\date{\today}

\maketitle

\section{Introduction}
The fact, that the spacetime could be noncommutative \cite{michael,bala1} at
very small length scale, received much attention in recent years.  The typical
length scale relevant for noncommutative physics is Planck scale. However, one
usually looks for phenomenological consequence of noncommutativity in quantum
mechanical regime. On the other hand, quantum mechanics is  a simple platform
for studying noncommutative physics and testing \cite{zhang1} the effect of
noncommtativity. In noncommutative spacetime the ordinary product is replaced
by a $star$ $product$ of the form
\begin{eqnarray}
\psi(x)\star\phi(x)=\psi(x)\exp\left\{-i\theta^{\mu\nu}\frac{\overleftarrow{\partial}}{\partial
x^\mu}\frac{\overrightarrow{\partial}}{\partial
y^\nu}\right\}\phi(y)\mid_{x=y}\,,\label{star1}
\end{eqnarray}
where the parameter $\theta^{\mu\nu}, \mu,\nu=1,2,3$,
is antisymmetric with respect to $\mu$
and $\nu$. It is easy to check that, (\ref{star1}) leads to the coordinate
noncommutativity $[x^\mu,x^\nu]= i\theta^{\mu\nu}$, by simply replacing
$\psi=x^\mu, \phi=x^\nu$ in (\ref{star1}). This type of  noncommutativity has
been used in various fields like quantum field theory, string theory, quantum
mechanics \cite{nair,pghosh1,Mikhail1}.

On the other hand in recent years there have been a great deal of interest in
the sudy of non Hermtian (as well as $\cal{PT}$ symmetric) quantum mechanics
\cite{bender1,bender2,bender3,bender4,bender5}. Recently there has also been
attempts to create $\cal{PT}$ symmetric systems in the laboratory
\cite{bender6}. Here our objective is to study non Hermitian quantum mechanics
in non commutative space. To be more specific we shall study models which are
$\cal{PT}$ symmetric as well as those which are non Hermitian and non
$\cal{PT}$ symmetric. It may be noted that noncommutative physics, defined by
$[x^\mu,x^\nu]= i\theta^{\mu\nu}$, is not $\mathcal{PT}$-symmetric.  We see
that  $\theta^{\mu\nu}$ does not change under $\mathcal{PT}$
transformation. The fact that $\theta^{\mu\nu}$  does not change under
$\mathcal{P}$ and  $\mathcal{T}$, hence under  $\mathcal{PT}$ has been pointed
out in  \cite{jabbari}.
However, in \cite{jabbari} certain transformation properties have been 
imposed so that the theory achieve a certain symmetry.

In this
article we shall first consider a $\cal{PT}$ symmetric deformation of non
commutative space and examine some models in this space. Secondly we shall
also examine non Hermitian (but not necessarily $\cal{PT}$ symmetric) models
defined in standard non commutative space. It will be seen that in quite a few
cases it is possible to obtain real spectrum despite the models being non
Hermitian. We consider our problem on a plane. The  $x_3$ coordinate commutes
with the coordinates of the plane and the dynamics along the $x_3$ direction 
is taken to free throughout our discussion.

The remainder of this article is organized as follows: In the next section we
propose a $\mathcal{PT}$-symmetric generalization of the noncommutative
quantum mechanics. As a simple example we then discuss the isotropic
oscillator in this $\mathcal{PT}$-symmetric noncommutative space. In section
\ref{ptoscillator} we discuss a $\mathcal{PT}$-symmetric displaced oscillator
in our formulation of noncommutative space and solve it. Here we also discuss
the same model in the standard non commutative space. A non-hermitian
oscillator is then discussed in \ref{ani}. Finally,  section \ref{conclu} is
devoted to a conclusion.

\section{$\mathcal{PT}$-symmetric deformation}\label{s3}
In this section we discuss the noncommutative quantum mechanics (QM), where
the noncommutativity respects $\mathcal{PT}$ symmetry, unlike the the standard
formalism, where the noncommutative algebra \cite{bertolami}
\begin{eqnarray}
\nonumber \left[\hat{x_\mu},\hat{x_\nu}\right]= i2\theta_{\mu\nu},~
\left[\hat{p_\mu},\hat{p_\nu}\right]=i2\hat{\theta_{\mu\nu}}\,,\\
\left[\hat{x_\mu},\hat{p_\nu}\right]= i\hbar(\delta_{\mu\nu}
+\theta_{\mu}^\beta\hat{\theta_{\nu\beta}})\,, \label{al4}
\end{eqnarray}
does not respect $\mathcal{PT}$-symmetry. Note that $\mathcal{PT}$ symmetry is
broken by the first two relations of the algebra (\ref{al4}) but the third one
respects the symmetry
\begin{eqnarray}
\nonumber\mathcal{PT}\left[\hat{x_\mu},\hat{x_\nu}\right]\mathcal{PT}^{-1}&\neq&
\mathcal{PT}i2\theta_{\mu\nu}\mathcal{PT}^{-1}\,,\\ \nonumber
\mathcal{PT}\left[\hat{p_\mu},\hat{p_\nu}\right]\mathcal{PT}^{-1}&\neq&
\mathcal{PT}i2\hat{\theta_{\mu\nu}}\mathcal{PT}^{-1},\\
\mathcal{PT}\left[\hat{x_\mu},\hat{p_\nu}\right]\mathcal{PT}^{-1}&=&
\mathcal{PT}i\hbar(\delta_{\mu\nu}
+\theta_{\mu}^\beta\hat{\theta_{\nu\beta}})\mathcal{PT}^{-1}\,.
\label{al4pt}
\end{eqnarray}
It is also clear from the representation
\begin{eqnarray}
\hat{x_\mu}=x_\mu -\theta_{\mu\nu} p^\nu,~~ \hat{p_\mu}=p_\mu
+\hat{\theta_{\mu\nu}} x^\nu\,,
 \label{rep3}
\end{eqnarray}
of the noncommutative coordinates on the phase space that they do not behave
as the commutative coordinates
\begin{eqnarray}
\mathcal{PT}\hat{x_\mu}\mathcal{PT}^{-1}\neq -\hat{x_\mu},~~
\mathcal{PT}\hat{p_\mu}\mathcal{PT}^{-1}\neq \hat{p_\mu}\,,
 \label{rep3npt}
\end{eqnarray}
under $\mathcal{PT}$ transformation.  We can however consider a noncommutative
formulation where the commutator algebra are $\mathcal{PT}$-symmetric. This
can be done by replacing $\theta_{\mu\nu}\to i\theta_{\mu\nu}$,
$\hat{\theta_{\mu\nu}}\to i\hat{\theta_{\mu\nu}}$ in the algebra (\ref{al4})
\begin{eqnarray}
\nonumber \left[\hat{x_\mu},\hat{x_\nu}\right]= -2\theta_{\mu\nu},~
\left[\hat{p_\mu},\hat{p_\nu}\right]=-2\hat{\theta_{\mu\nu}}\,,\\
\left[\hat{x_\mu},\hat{p_\nu}\right]= i\hbar(\delta_{\mu\nu}
-\theta_{\mu}^\beta\hat{\theta_{\nu\beta}})\,, \label{al4pt}
\end{eqnarray}
It is useful to consider the same representation (\ref{rep3}) but with the
replacement $\theta_{\mu\nu}\to i\theta_{\mu\nu}$, $\hat{\theta_{\mu\nu}}\to
i\hat{\theta_{\mu\nu}}$ as
\begin{eqnarray}
\hat{x_\mu}&=&x_\mu -i\theta_{\mu\nu} p^\nu,~~ \hat{p_\mu}=p_\mu
+i\hat{\theta_{\mu\nu}} x^\nu \,,
 \label{rep3pt}
\end{eqnarray}
which  transforms in the same way as the ordinary coordinate and
momentum under $\mathcal{PT}$-transformation. It is now obvious that
any quantum mechanical system which is $\mathcal{PT}$-symmetric in
commutative space will remain $\mathcal{PT}$-symmetric when
considered in the noncommutative space given by the algebra
(\ref{al4pt}). To discuss the QM in this formulation we consider the
isotropic oscillator on a plane.
\begin{eqnarray}
H = \frac{1}{2m}(p_1^2+p_2^2)+ \frac{1}{2}m\omega^2(x_1^2+x_2^2)\,,
\label{2dh}
\end{eqnarray}
Eq. (\ref{2dh}) is both Hermitian and
$\mathcal{PT}$-symmetric. The corresponding noncommutative
counterpart
\begin{eqnarray}
\hat{H} = \frac{1}{2m}(\hat{p_1}^2 +\hat{p_2}^2)+
\frac{1}{2}m\omega^2(\hat{x_1}^2+\hat{x_2}^2)\,, \label{2dhn}
\end{eqnarray}
then takes the form
\begin{eqnarray}
\nonumber H_\mathcal{PT} &=& \frac{1}{2M_\mathcal{PT}}(p_1^2+p_2^2)+
\frac{1}{2}M_\mathcal{PT}\Omega_{\mathcal{PT}}^2(x_1^1+x_2^2)\\& -&
iS_{\mathcal{PT}}L_z\,, \label{2dh1pt}
\end{eqnarray}
where  $1/M_\mathcal{PT}= 1/m -m\omega^2\theta^2$,
$S_\mathcal{PT}=(m\omega^2\theta +\hat{\theta}/m)$, and $\Omega_\mathcal{PT}=
\sqrt{(1/m -m\omega^2\theta^2)(m\omega^2 -\hat{\theta}^2/m)}$. Note that, in
order to get (\ref{2dh1pt}) we used the  relation (\ref{rep3pt}) with
$\theta_{12}=\theta$, $\hat{\theta_{12}}= \hat\theta$ and $\mu,\nu=1,2$. It is
possible to evaluate the spectrum of the Hamiltonian (\ref{2dh1pt}) for a
special case $S_\mathcal{PT}=0$, which makes it isotropic. The eigenvalues are

\begin{eqnarray}
E_\mathcal{PT}= \Omega_\mathcal{PT}\left(n^++ n^- +1\right)\,,
\label{eigenpt}
\end{eqnarray}
where $n^\pm \in \mathbb{N}$, $\Omega_\mathcal{PT}=
\omega(1-m^2\omega^2\theta^2)= \omega(1+\theta\hat\theta)$. However for
general case, $S_\mathcal{PT}\neq 0$, although the Hamiltonian (\ref{2dh1pt})
is $\mathcal{PT}$-symmetric the solution of the corresponding eigenvalue
problem can be complex.  It is possible to write (\ref{2dh1pt}) as a sum of
two separate one dimensional harmonic oscillators with frequencies
$\Omega_{\mathcal{PT}} + iS_{\mathcal{PT}}$ and
$\Omega_{\mathcal{PT}}-iS_{\mathcal{PT}}$,
\begin{eqnarray}
\nonumber H_{\mathcal{PT}}=
\Omega_{\mathcal{PT}}\left(a_{\mathcal{PT}}^\dagger a_{\mathcal{PT}} +
b_{\mathcal{PT}}^\dagger b_{\mathcal{PT}} +1\right) \\+
iS_{\mathcal{PT}}\left(a_{\mathcal{PT}}^\dagger a_{\mathcal{PT}}
-b_{\mathcal{PT}}^\dagger b_{\mathcal{PT}}\right)\,. \label{2dosh}
\end{eqnarray}
Here the annihilation operators $a_{\mathcal{PT}}$ and $b_{\mathcal{PT}}$ are
\begin{eqnarray}
a_{\mathcal{PT}}=\frac{\left[(p_1+ip_2) -
iM_{\mathcal{PT}}\Omega_{\mathcal{PT}}(x_1+ix_2)
\right]}{2\sqrt{M_\mathcal{PT}\Omega_{\mathcal{PT}}}}\,,\\
b_{\mathcal{PT}}=\frac{\left[(p_1-ip_2) -
iM_{\mathcal{PT}}\Omega_{\mathcal{PT}}(x_1-ix_2)
\right]}{2\sqrt{M_{\mathcal{PT}}\Omega_{\mathcal{PT}}}}\,.
\label{annihi1}
\end{eqnarray}
The corresponding creation operators $a_{\mathcal{PT}}^\dagger,
b_{\mathcal{PT}}^\dagger$ together with the annihilation operators satisfy the
usual commutation relations ($\hbar=1$ unit is used)
\begin{eqnarray}
\left[a_{\mathcal{PT}},
a_{\mathcal{PT}}^\dagger\right]=\left[b_{\mathcal{PT}},
b_{\mathcal{PT}}^\dagger\right] =1\,,\label{2doshac}
\end{eqnarray}
all other commutators are zero.  The number operators
$\mathcal{N}_{\mathcal{PT}}^+=a_{\mathcal{PT}}^\dagger a_{\mathcal{PT}}$ and
$\mathcal{N}_{\mathcal{PT}}^-=b_{\mathcal{PT}}^\dagger b_{\mathcal{PT}}$
satisfy the eigenvalue equation $\mathcal{N}_{\mathcal{PT}}^\pm|n^+,n^-\rangle
= n^\pm|n^+,n^-\rangle$ with $n^\pm \in \mathbb{N}_0$. The exact eigenvalue
for the Hamiltonian (\ref{2dosh}) can be found as
\begin{eqnarray}
E_{\mathcal{PT}}= \Omega_{\mathcal{PT}}(n^++ n^- +1) +iS_{\mathcal{PT}}(n^+-
n^-)\,. \label{2dosei}
\end{eqnarray}
Note that the eigenvalue is real only for $n^+= n^-= n$. In that case
$E_{\mathcal{PT}}= \Omega_{\mathcal{PT}}(2n +1)$.

It would be informative at this point to mention the result of isotropic
oscillator (\ref{2dh}) studied on the  standard noncommutative space
\cite{agni} defined  by (\ref{al4}). The $2$-dimensional isotropic oscillator
on a plane with only noncommutative coordinates becomes anisotropic with two
different frequencies. Even for both the coordinates and momenta  to be
noncommutative like (\ref{al4}), the oscillator becomes anisotropic with
completely real eigenvalues \cite{giri1}
\begin{eqnarray}
E_{NC} = \Omega_{NC}(n^++ n^- +1) +S_{NC}(n^+- n^-)\,. \label{2dosein}
\end{eqnarray}
where, $\Omega_{NC}= \sqrt{(1/m +m\omega^2\theta^2)(m\omega^2
+\hat{\theta}^2/m)}$ and $S_{NC}=(m\omega^2\theta +\hat{\theta}/m)$ and
$1/M_{NC}= 1/m +m\omega^2\theta^2$.  To obtain the above result for $E_{NC}$,
the relation (\ref{rep3}), with $\theta_{12}=\theta$,
$\hat{\theta_{12}}=\hat\theta$ and $\mu,\nu=1,2$ has been used.

\section{$\mathcal{PT}$-symmetric oscillator}\label{ptoscillator}
In this section, we consider a $\mathcal{PT}$-symmetric oscillator given by
the Hamiltonian
\begin{eqnarray}
\nonumber H_{\mathcal{PT}} &=& \frac{1}{2m}\left(p_1^2+p_2^2\right)+
\frac{1}{2}m\omega^2\left(x_1^2+x_2^2\right)\\ &+&iC
\left(x_1+x_2\right)\,. \label{pt2dh}
\end{eqnarray}
where the real constant $C$ is the strength of $\mathcal{PT}$-symmetric
interaction. It has real eigenvalues \cite{bender1} with the ground state
shifted. The noncommutative version  defined as
\begin{eqnarray}
\nonumber \hat{H_{\mathcal{PT}}} &=&
\frac{1}{2m}\left(\hat{p_1}^2+\hat{p_2}^2\right)+
\frac{1}{2}m\omega^2\left(\hat{x_1}^2+\hat{x_2}^2\right)\\ &+&iC
\left(\hat{x_1}+\hat{x_2}\right)\,, \label{pt2dhn}
\end{eqnarray}
is $\mathcal{PT}$-symmetric,
$\mathcal{PT}\hat{H_{\mathcal{PT}}}\mathcal{PT}^{-1}=\hat{H_{\mathcal{PT}}}$.
Note that here we used the algebra (\ref{al4pt}) and the representation
(\ref{rep3pt}). In terms of  commutative coordinates (\ref{pt2dhn}) can be
written as
\begin{eqnarray}
\nonumber \hat{H_\mathcal{PT}} &=&
\frac{1}{2M_{\mathcal{PT}}}\left(p_1^2+p_2^2\right)+
\frac{1}{2}M_{\mathcal{PT}}\Omega_{\mathcal{PT}}^2\left(x_1^2+x_2^2\right)
\\&-& iS_{\mathcal{PT}}L_z
+iC\left(x_1+x_2\right)-C\theta\left(p_1-p_2\right)\,.
\label{pt2dh1pt}
\end{eqnarray}
We first solve the problem (\ref{pt2dh1pt}) for a special case
$S_{\mathcal{PT}}=0$.  We use the following  transformation
\begin{eqnarray}
\nonumber Q_1 &=&x_1 + \frac{iC}{M_{\mathcal{PT}}\Omega_{\mathcal{PT}}^2},~~
Q_2=x_2 + \frac{iC}{M_{\mathcal{PT}}\Omega_{\mathcal{PT}}^2}\\ P_1 &=&
p_1-C\theta M_{\mathcal{PT}},~~ P_2=p_2+ C\theta M_{\mathcal{PT}}\label{trans}
\end{eqnarray}
on (\ref{pt2dh1pt}), which  then becomes an isotropic oscillator with a
constant shift,
\begin{eqnarray}
\hat{H_\mathcal{PT}} = \frac{1}{2M_{\mathcal{PT}}}\left(P_1^2+P_2^2\right)+
\frac{1}{2}M_{\mathcal{PT}}\Omega_{\mathcal{PT}}^2\left(Q_1^2+Q_2^2\right)
+E_0\,, \label{pt2dh1pt2}
\end{eqnarray}
where the constant is $E_0=
\frac{C^2}{M_{\mathcal{PT}}\Omega^2_{\mathcal{PT}}}-
C^2\theta^2M_{\mathcal{PT}}$. The eigenvalue is readily obtained to be
\begin{eqnarray}
\hat{E_\mathcal{PT}}= \Omega_\mathcal{PT}\left(n^++ n^- +1\right) +E_0\,,
\label{eigenpt2}
\end{eqnarray}
However for general case, $S_\mathcal{PT}\neq 0$, although the Hamiltonian
(\ref{pt2dh1pt}) is $\mathcal{PT}$-symmetric, the solution of the
corresponding eigenvalue problem can be complex. To solve the eigenvalue
problem we make the following transformation
\begin{eqnarray}
Q_1 &=&x_1 + i\lambda\,, Q_2 =x_2 + i\lambda\,,\\ P_1 &=& p_1+\beta \,, P_2 =
p_2- \beta\,,\label{trans1}
\end{eqnarray}
where $\lambda=\frac{C(\theta M_{\mathcal{PT}}S_{\mathcal{PT}}
+1)}{M_{\mathcal{PT}}( \Omega_{\mathcal{PT}}^2 + S^2_{\mathcal{PT}})}$ and
$\beta=\frac{C(S_{\mathcal{PT}}-\theta
M_{\mathcal{PT}}\Omega^2_{\mathcal{PT}})}{S^2_{\mathcal{PT}}+\Omega^2_{\mathcal{PT}}}
$. After this transformation Eq. (\ref{pt2dh1pt}) takes the form
\begin{eqnarray}
\nonumber \hat{H_\mathcal{PT}} &=&
\frac{1}{2M_{\mathcal{PT}}}\left(P_1^2+P_2^2\right)+
\frac{1}{2}M_{\mathcal{PT}}\Omega_{\mathcal{PT}}^2\left(Q_1^2+Q_2^2\right)\\&-&iS_{\mathcal{PT}}(Q_1P_2-Q_2P_1)
+\hat{E_0}\,, \label{pt2dh1pt2}
\end{eqnarray}
where $\hat{E_0}= \beta^2/M_{\mathcal PT} - M_{\mathcal PT}\Omega^2_{\mathcal
PT}\lambda^2 -2S_{\mathcal PT}\lambda\beta + 2C\lambda +2C\theta\beta$. The
exact eigenvalues for the Hamiltonian (\ref{pt2dh1pt}) can be found as
\begin{eqnarray}
\hat{E_{\mathcal{PT}}}= \Omega_{\mathcal{PT}}(n^++ n^- +1)
+iS_{\mathcal{PT}}(n^+- n^-) +\hat{E_0}\,. \label{2dosei1}
\end{eqnarray}
Note that the eigenvalues are real only for $n^+= n^-= n$, in that case
$\hat{E_{\mathcal{PT}}}= \Omega_{\mathcal{PT}}(2n +1) + \hat{E_0}$. One can
see that the limit $\hat{\theta} \to 0$ of Eq.  (\ref{2dosei1}) can be taken
smoothly.

Note that although (\ref{2dosei1}) respects $\mathcal{PT}$-symmetry, it
becomes complex in general. It would now be interesting to see what happens if
we consider the standard noncommutativity, given by (\ref{al4}), to study the
$\mathcal{PT}$-symmetric Hamiltonian (\ref{pt2dh}). This can be simply
achieved by replacing $i\theta \to \theta$ and $i\hat\theta \to \hat\theta$
from Eq. (\ref{pt2dh1pt}) onwards. It may be seen that in this case the
Hamiltonian (\ref{pt2dh}) is non Hermitian and non $\cal{PT}$ symmetric. The
eigenvalue are then given by
\begin{eqnarray}
\hat{E_{NC}}= \Omega_{NC}(n^++ n^- +1) + S_{NC}(n^+- n^-)
+\epsilon_{NC}\,. \label{NCeigen1}
\end{eqnarray}
where, $\epsilon_{NC}$
\begin{eqnarray}
\nonumber &=& -\frac{C^2(S_{NC}-\theta
M_{NC}\Omega^2_{NC})^2}{(-S^2_{NC}+\Omega^2_{NC})^2M_{NC}} -
\frac{C^2\Omega_{NC}^2(1-\theta M_{NC}S_{NC})^2}{M_{NC}( \Omega_{NC}^2
-S^2_{NC})^2}\\ \nonumber & +&\frac{2C^2S_{NC}(1-\theta
M_{NC}S_{NC})(S_{NC}-\theta
M_{NC}\Omega_{NC})}{M_{NC}(\Omega^2_{NC}-S_{NC}^2)} \\ \nonumber &+&
\frac{2C^2(1-\theta M_{NC}S_{NC})}{M_{NC}(\Omega^2_{NC}-S^2_{NC})}
-\frac{2C^2\theta(S_{NC}-\theta M_{NC}\Omega^2_{NC})}
{(\Omega^2_{NC}-S^2_{NC})}
\end{eqnarray}
Note that, this time the eigenvalues (\ref{NCeigen1}) are completely real.

\section{Anisotropic oscillator with non-hermitian coupling}\label{ani}
We consider anisotropic oscillator on a plane with a non-hermitian coupling of
the form $\sim ix_1x_2$. The Hamiltonian  we consider is of the form
\cite{asiri}
\begin{eqnarray}
H=  \frac{1}{2}\sum_{i=1}^{2}\left(p_i^2 + C_ix_i^2\right) +
\frac{1}{2}iC_3x_1x_2\label{ham}
\end{eqnarray}
The Hamiltonian (\ref{ham}) is not hermitian $H^\dagger \neq H$. We now
consider this Hamiltonian in noncommutative space. It is however better to
follow the transformation of Ref. \cite{asiri}, which is of the form
\begin{eqnarray}
\nonumber X &=& \alpha_1x_1 + x_2\,,~~~ Y= \alpha_2x_1 + x_2\,,\\ P_X &=&
\frac{p_1-\alpha_2p_2}{\alpha_1-\alpha_2}\,,~P_Y=
\frac{p_1-\alpha_1p_2}{\alpha_2-\alpha_1}\,.  \label{trans}
\end{eqnarray}
where $\alpha_1=\left((C_1^2-C_2^2)-\sqrt{(C_1^2-C_2^2)^2-
C_3^2}\right)/(iC_3)$ and $\alpha_1=\left((C_1^2-C_2^2)+\sqrt{(C_1^2-C_2^2)^2-
C_3^2}\right)/(iC_3)$.  Note that the new variables satisfy the usual
commutation rules $[X,P_X]= [Y,P_Y]= i\hbar$. All other commutators being
zero. In terms of the new canonical variables $X, Y, P_X, P_Y$, the
Hamiltonian (\ref{ham}) reads as
\begin{eqnarray}
H=  e^{K}H_K + e^{-K}H_{-K}\,,\label{ham12}
\end{eqnarray}
with the two separate one dimensional Hamiltonians
\begin{eqnarray}
H_K= \frac{\tilde{P_X}^2}{2m} + \frac{1}{2}m\Omega^2\tilde{X}^2\,, H_{-K}=
\frac{\tilde{P_Y}^2}{2m} + \frac{1}{2}m\Omega^2\tilde{Y}^2\,, \label{ham123}
\end{eqnarray}
where the new canonical phase space coordinates are
\begin{eqnarray}
\nonumber \tilde{P_X} &=&
e^{-K/2}\sqrt[4]{\frac{1+\alpha_1^2}{1+\alpha_2^2}}P_X\,, \tilde{P_Y}=
e^{K/2}\sqrt[4]{\frac{1+\alpha_2^2}{1+\alpha_1^2}}P_Y,\\ \tilde{X} &=&
e^{K/2}\sqrt[4]{\frac{1+\alpha_2^2}{1+\alpha_1^2}}X\,, \tilde{Y}=
e^{-K/2}\sqrt[4]{\frac{1+\alpha_1^2}{1+\alpha_2^2}}Y\,.
\label{newcoor}
\end{eqnarray}
The two oscillators of (\ref{ham12}) share the same mass  $m=
1/\sqrt{(1+\alpha_1^2)(1+\alpha_2^2)}$, and have the frequencies
\begin{eqnarray}
\nonumber \Omega
e^K=\sqrt{1/2\left[(C_1^2+C_2^2)-\sqrt{(C_1^2-C_2^2)^2-C_3^2}\right]} ,\\
\Omega
e^{-K}=\sqrt{1/2\left[(C_1^2+C_2^2)+\sqrt{(C_1^2-C_2^2)^2-C_3^2}\right]}\,.
\label{freq11}
\end{eqnarray}
The explicit form of the parameters $\Omega$ and $K$ can be found from
(\ref{freq11}) as $\Omega=\sqrt[4]{( C_1^2C_2^2+ C_3^2/4)}$ and
$e^K=\sqrt{\frac{\left[(C_1^2+C_2^2)-\sqrt{(C_1^2-C_2^2)^2-C_3^2}\right]}{
\left[(C_1^2+C_2^2)+\sqrt{(C_1^2-C_2^2)^2-C_3^2}\right]}}$. Note that  in
oreder to keep two frequencies of (\ref{freq11}) real,  the condition
$|C_1^2-C_2^2| \geq C_3$ needs to be satisfied.  Thus (\ref{ham12}) is an
anisotropic oscillator and its eigenvalues are \cite{asiri}
\begin{eqnarray}
E=\Omega\left[e^{K}(n_K +1/2) + e^{-K} (n_{-K} +1/2)\right]\,,
\label{eigen123}
\end{eqnarray}
where $n_K, n_{-K}\in \mathbb{N}$. Before discussing the
$\mathcal{PT}$-symmetric noncommutative quantum mechanics for (\ref{ham12}),
we first discuss the problem in standard noncommutative space. We consider
only coordinates to be noncommutative of the form
\begin{eqnarray}
\left[\hat{\tilde{X}},\hat{\tilde{Y}}\right]=
2i\underline\theta\,,\label{nonX1}
\end{eqnarray}
with the representation
$\hat{\tilde{X}}=\tilde{X}-\underline\theta\tilde{P_Y},
\hat{\tilde{Y}}=\tilde{Y}+\underline\theta\tilde{P_X}$. Note that consistency
with (\ref{al4}) and (\ref{rep3}) demands that $\underline\theta=
(\alpha_1-\alpha_2)\theta$. The noncommutative version of (\ref{ham12}) is
\begin{eqnarray}
\nonumber H &=&  e^{K}H_K + e^{-K}H_{-K} \\ \nonumber &+&
e^K\left(m^2\Omega^2\underline\theta^2\tilde{P_Y}^2-m\Omega^2\underline\theta
\tilde{X}\tilde{P_Y}\right)\\
&+&e^{-K}\left(m^2\Omega^2\underline\theta^2\tilde{P_X}^2+m\Omega^2\underline\theta
\tilde{Y}\tilde{P_X}\right)\label{ham12N}
\end{eqnarray}
The perturbative spectrum of (\ref{ham12N}) is given by
\begin{eqnarray}
\nonumber E &=&\Omega\left[e^{K}(n_K +1/2) + e^{-K} (n_{-K} +1/2)\right]\\
&+&\Omega^3 m^3\underline{\theta}^2\left[e^{K}(n_K +1/2) + e^{-K} (n_{-K}
+1/2)\right]. \label{eigen123p1}
\end{eqnarray}

Now let us consider the $\mathcal{PT}$-symmetric non-commutative case.  In
noncommutative phase space, the coordinates $X, Y, P_X, P_Y$ of (\ref{ham12})
are replaced by the corresponding noncommutative counterparts $\hat{X},
\hat{Y}, \hat{P_X}, \hat{P_Y}$, which lead to
\begin{eqnarray}
\nonumber \hat{H}=  e^{K}H_K + e^{-K}H_{-K} \hspace{5cm}\\ \nonumber -
e^{K}\left(\frac{\underline{\hat\theta}^2}{2m} \tilde{Y}^2
+\frac{m\Omega^2\underline{\theta}^2}{2}\tilde{P_Y}^2
+i{\underline{\hat\theta}}{m}\tilde{Y}\tilde{P_X}-i
m\Omega^2\underline{\theta}\tilde{X}\tilde{P_Y}\right) \\-
e^{-K}\left(\frac{\underline{\hat\theta}^2}{2m} \tilde{X}^2
+\frac{m\Omega^2\underline{\theta}^2}{2}\tilde{P_X}^2
-i{\underline{\hat\theta}}{m}\tilde{X}\tilde{P_Y}+i
m\Omega^2\underline{\theta}\tilde{Y}\tilde{P_X}\right)
 \label{ham1hat}
\end{eqnarray}
Note that the noncommutative operators $\hat{X}, \hat{Y}, \hat{P_X},
\hat{P_Y}$, now satisfy the commutation algebra
\begin{eqnarray}
\nonumber \left[\hat{X},\hat{Y}\right]&=& -2\underline{\theta}\,,
\left[\hat{P_X},\hat{P_Y}\right]= -2\underline{\hat\theta}\,,\\
\left[\hat{X},\hat{P_X}\right]&=& \left[\hat{P},\hat{P_Y}\right]=
i\hat\hbar\,,\label{communew}
\end{eqnarray}
where $\underline{\theta}= (\alpha_1-\alpha_2)\theta$,
$\underline{\hat\theta}= \hat\theta/(\alpha_1-\alpha_2)$ and  the
representation
\begin{eqnarray}
\nonumber \hat{X} &=& X +i\underline{\theta}P_Y\,, \hat{Y}= Y
-i\underline{\theta}P_X\\ \hat{P_X}&=& P_X -i\underline{\hat\theta}Y\,,
\hat{P_Y}= P_Y +i\underline{\hat\theta}X\,, \label{repX}
\end{eqnarray}
which can be obtained from (\ref{rep3pt}), is used to get (\ref{ham1hat}). It
is possible to solve (\ref{ham1hat}) by perturbation. The eigenvalues are
given by
\begin{eqnarray}
\nonumber E &=&\Omega\left[e^{K}(n_K +1/2) + e^{-K} (n_{-K} +1/2)\right]\\
\nonumber &+&\frac{1}{2}\Omega^3 m^2\underline{\theta}^2\left[e^{K}(n_K +1/2)
+ e^{-K} (n_{-K} +1/2)\right]\\
&+&\frac{1}{2}\frac{\hat{\underline{\theta}}^2}{m^2\Omega}\left[e^{K}(n_K
+1/2) + e^{-K} (n_{-K} +1/2)\right]\,.
 \label{eigen123p2}
\end{eqnarray}
Note that all the complex terms of (\ref{ham1hat}) has the zero expectation
values, which makes the spectrum (\ref{eigen123p2}) real. Importantly,
as mentioned
before, the condition for reality of frequencies, $|C_1^2-C_2^2| \geq C_3$,
must have to be satisfied by (\ref{eigen123p1}) and (\ref{eigen123p2}) in
order to keep the spectrum real. Othewise, it will generate complex
eigenvalues in conjugate pairs.


\section{Conclusion and discussion}\label{conclu}
In this article $\mathcal{PT}$-symmetric quantum mechanical model is studied
in noncommutative space. In order to keep the $\mathcal{PT}$-symmetry intact,
we needed a non-commutative formulation which itself respects
$\mathcal{PT}$-symmetry. This is done by replacing the non-commutativity
parameter with pure imaginary parameter. However the systems on the
$\mathcal{PT}$-symmetric non-commutative spaces generates complex eigen-values
in some cases. On the other hand non Hermitian systems in standard
noncommutative space has been found to have entirely real spectrum.

\end{document}